\documentstyle[12pt]{article}
\newcommand{\be}{\begin{equation}}
\newcommand{\ee}{\end{equation}}
\newcommand{\ba}{\begin{eqnarray}}
\newcommand{\ea}{\end{eqnarray}}

\begin{document}
\begin{center}
 {\bf\Large{
  Chain and Hamilton - Jacobi approaches
 for systems with purely second class constraints}}
\end{center}
\begin{center} {\bf Dumitru Baleanu}\footnote[1]{ On leave of absence from
Institute of Space Sciences, P.O BOX, MG-23, R 76900
Magurele-Bucharest, Romania, E-mail: dumitru@cankaya.edu.tr} and
{\bf Yurdahan
G\"uler}\footnote[2]{E-Mail:~~yurdahan@cankaya.edu.tr}
\end{center}
\begin{center}
Department of Mathematics and Computer Sciences, Faculty of Arts
and Sciences, Cankaya University-06530, Ankara , Turkey
\end{center}
\begin{abstract}
 The equivalence of the chain method and Hamilton-Jacobi formalism
 is demonstrated. The stabilization algorithm of Hamilton-Jacobi formalism is clariffied and
 two examples are presented in details.
\end{abstract}
\newpage
 \section {Introduction}

    The quantization of second-class constrained systems, initiated by
   Dirac\\
\cite{dirac} and developed by various authors \cite{hanson}
  is subjected to intense debates during
the last years. A powerful method, based on the introduction of
extra variables was proposed in \cite{batalin} and it leads us to
convert the second class-constraints into first class ones.
Recently, in \cite{marnelius} a second-class constrained system
was investigated and  a gauge theory was found without introducing
extra variables. In \cite{mitra} the authors proposed to eliminate
half of the second-class constraints and converted the second half
to the first-class constraints. This method is called the
chain-method and it was subjected recently to various
investigations \cite{mojiri}
  In the chain method one primary constraint is producing, under some
assumptions,  a set of constraints.At the end all primary
constraints produce a chain of second class constraints but some
of them are in involution. The next step of the method is to
produce a gauge theory by adding an extra term in the canonical
Hamiltonian in such a manner that half of the constraints are
eliminated \cite{mitra}.Despite of many attempts to elucidate the
integrability
 conditions of the second-class constrained systems in Hamilton-
 Jacobi formalism (HJ) based on $Carath{\acute e}odory$'s approach
\cite{cara} and initiated in \cite{gu}
 there are some hidden parts which must be clarified.
The main problem is to clarify when the stabilization algorithm
come to an end and to make a deep exploration of the equivalence
with Dirac's formulation. On the other hand we have to keep in
mind the physical interpretation of (HJ) partial differential
equations involved in (HJ) formalism \cite{gol}. All this issues
are important for the construction of the integrability of the
second-class constrained systems in (HJ) formalism \cite{bal}.

  The main aim of this paper is to show that the stabilization algorithms produce the same constraints
  and that the lagrange multipliers of chain method
  arise naturally in (HJ) formalism.

  The plan of the paper is as follows:

  In sec. 2 the (HJ) formalism and the chain method are briefly
 presented.
In sect. 3 the equivalence of both method are investigated and two
examples were analyzed. Sec. 4 is devoted to conclusions.

\section{The methods}

\subsection{Hamilton-Jacobi formalism}

Let us assume that the Lagrangian L is singular and the Hessian
supermatrix has rank n-r. The Hamiltonians to start with are

\be\label{doi} H_{\alpha}^{'}=H_{\alpha}(t_{\beta},q_{a},p_{a})
 +p_{\alpha}, \ee where $\alpha,\beta=n-r +1,\cdots,n$,$a=1,\cdots
n-r$. The usual Hamiltonian $H_0$ is defined as

\be\label{unu} H_{0}=-L(t,q_{i},{\dot q_{\nu}},{\dot q_{a}=w_{a}})
+p_{a}w_{a} + {\dot
q_{\mu}}p_{\mu}\mid_{p_{\nu}=-H_{\nu}},\nu=0,n-r+1,\cdots,n. \ee
which is independent of $\dot q_{\mu}$.Here $\dot
q_{a}={dq_{a}\over d\tau}$,where $\tau$ is a parameter.
 The equations of motion are obtained as total differential equations
in many variables as follows

\ba\label{(pq)} &dq_{a}&=(-1)^{P_a +P_a P_\alpha} {\partial_{r}
H_{\alpha}^{'}\over\partial p_{a}}dt_{\alpha},
dp_{a}=-(-1)^{P_{a}P_{\alpha}}{\partial_{r}
H_{\alpha}^{'}\over\partial q_{a}}dt_{\alpha},\cr
&dp_{\mu}&=-(-1)^{P_{\mu}P_{\alpha}}{\partial_{r}
H_{\alpha}^{'}\over\partial t_{\mu}}dt_{\alpha}, \mu=1,\cdots, r ,
\ea

\be\label{(z)} dz=(-H_{\alpha} + (-1)^{P_a + P_a
P_\alpha}p_{a}{\partial_{r} H_{\alpha}^{'}\over\partial
p_{a}})dt_{\alpha}, \ee where $z=S(t_{\alpha},q_{a})$ and $P_{i}$
represents the parity  of $a_{i}$.

On the surface of constraints the system of differential
equations(\ref{(pq)}) is integrable if and only if
\be\label{condh} [H_{\alpha}^{'},H_{\beta}^{'}]=0, \forall \alpha
,\beta.\ee

 \subsection{The chain method}

 Let us consider a singular Lagrangian $L(q_a,{\dot q_a}),
a=1,\cdots N$ admitting a primary constraint \be\label{ecu11}
\phi_{1}(p_a,q_a)=0. \ee We assume that the Lagrangian possesses
only $2n(n<N)$ second-class constraints. The total Hamiltonian is
\be\label{totalh} H=H_c+\lambda\phi_1, \ee where $H_c$ is the
canonical Hamiltonian and $\lambda$ is a parameter to be
determined. Taking into account the integrability conditions  we
obtain the following one-chain system: \be\label{lant}
\phi_2(p_a,q_a)\equiv\{\phi_1,H\}=0,\cdots,\phi_{2n}(p_a,q_a)\equiv\{\phi_{2n-1},H\}=0.
\ee Since the set of constraints $\phi_i$ is composed of
second-class constraints, the matrix $ M_{ij}= \{\phi_i,\phi_j\}$
is nonsingular. Explicitly, \be\label{matrice}
M_{ij}=\left(\begin{array}{ccccccc}
0 & 0 & 0 & \cdots & 0 & 0 &  -\alpha\\
0 & 0 & 0 & \cdots & 0 & \alpha & *    \\
0 & 0 & 0 &         & -\alpha& * & *  \\
\vdots&  &         &        &  &  &\vdots \\
0 & 0 & \alpha     &  & 0 & *&*    \\
0 & -\alpha& * &   &        * & 0 & * \\
\alpha&*   & * &\cdots&*      &*& 0\\
\end{array}\right),on\quad \Sigma_{2n}.
\ee Here "on $\sum_{2n}$" means the weak equality modulo the
constraints $\phi_1=\phi_2=\cdots=\phi_{2n}=0$.
 On the other hand we have
 \be \{\phi_{1},\phi_{n}\}=-\alpha(p_a,q_a)
 \ee
 and half of the second class constraints, as we can be seen from (\ref{matrice}), are in involution
 \cite{mitra}.The advantage of this method is that it
gives some new information about the form of $M_{ij}$ and that
some of the constraints are in involution. The next step is to
generalize the procedure to a chain of K primary constraints.
  We denote these constraints by
  $\phi_1^{(1)},\phi_1^{(2)},\cdots,\phi_1^{(K)}$.
\be
\begin{array}{ccccccc}
\phi_1^{(1)}=0 &  & \phi_1^{(2)}=0 &  & \cdots &  & \phi_1^{(K)}=0 \\
\phi_2^{(1)}\equiv\{\phi_1^{(1)},H\}=0 &  & \phi_2^{(2)}\equiv\{\phi_1^{(2)},H\}=0 & & \cdots &  & \phi_2^{(K)}\equiv\{\phi_1^{(K)},H\}=0   \\
\phi_{M_1}^{(1)}\equiv\{\phi_{M_1-1}^{(1)},H\}=0 &  & \phi_{M_2}^{(2)}\equiv\{\phi_{M_2-1}^{(2)},H\}=0 & & \cdots &  & \phi_{M_K}^{(K)}\equiv\{\phi_K^{(K)},H\}=0   \\
\end{array}
\ee The Hamiltonian H is given by \be
H=H_c+\sum_{m=1}^{K}v_0^{(m)}\phi_{1}^{(m)} \ee

\be
 \begin{array}{ccccc}
 \{\phi_{i}^{(l)},\phi_{j}^{(m)}\}\approx0&, i\le r_l,& j\le r_m;l,m & arbitrary&,
 \end{array}
 \ee
Using the same arguments as before we observe that the $K\times K$
matrix \be
 \alpha\equiv\left(\begin{array}{ccccc}
 \alpha_{11}& \cdots & \alpha_{1K}& \\
 \vdots&     &          \vdots&\\
\alpha_{K1}& \cdots & \alpha_{KK}&
 \end{array}\right)
 \ee
 where $\alpha_{mn}=\{\phi_{r_l+1}^{(l)},\phi_{r_m}^{m}\}$ must
 have at least one non-zero element in each row \cite{mitra}
 The method is applicable to local field theory too. Let us assume
that the local Lagrangian density L is singular and admits only
one second-class constraints $\phi_{1}({\vec x}),\phi_{2}({\vec
x}),\cdots,\phi_{2n}({\vec x}) $ and only $\phi_{1}({\vec x})$ is
primary. In this method we have infinite number of constraint
equations. If we calculate the bracket of constraints we will
construct the matrix $M_{ij}=\{\phi_i,\phi_j\}$ as

\be\label{matrice2} M_{ij}({\vec x},{\vec
y})=\left(\begin{array}{ccccccc}
0 & 0 & 0 & \cdots & 0 & 0 &  -\alpha\\
0 & 0 & 0 & \cdots & 0 & \alpha & *    \\
0 & 0 & 0 &         & -\alpha& * & *  \\
\vdots&  &         &        &   &   &\vdots \\
0 & 0 & \alpha     &  & 0 & *&*    \\
0 & -\alpha& * &   &        * & 0 & * \\
\alpha&*   & * &\cdots&*      &*& 0\\
\end{array}\right)\delta({\vec x}-{\vec y})
\ee

To find the subset of constraints which are in involution we
follow the procedure used before.To simplify  the problem  and to
explain the method we consider the two-chain case. Let us assume
the set of the constraints in involution as
$\{\chi_1,\chi_2,\cdots,\chi_r;\psi_1,\psi_2,\cdots, \psi_{q}\}.$
To stop the chain we transform the Hamiltonian as \be\label{extra}
H^{''}=H_c+{1\over 2}\chi^{T}\alpha^{-1}\chi, \ee where

\be \chi=\left(\begin{array}{c} \chi_{r+1} \\
\psi_{q+1}\end{array}  \right) \ee

For field theory let us  assume that  only a primary constraint
$\Phi_1$ generates 2n-1 constraints denoted by $\Phi_{\alpha},
\alpha=2,\cdots 2n$. Using the same procedure as before we
construct the corresponding extended Hamiltonian as
\be\label{extra1} H^{'''}= H_c +{1\over 2}\int d{\vec
x}\alpha^{-1}({\vec x})\Phi_{n+1}^{2}(\vec x), \ee such that we
eliminated half of the constraints and the resulting system is a
first-class one.

\section{Equivalence of the methods}

Let us consider a singular Lagrangian L admitting 2n second class
constraints and K of them are primary. Let us denote the primary
constraints by $H_1^{'}, H_2^{'}\cdots, H_K^{'}$. Consistency
conditions read as
  \be\label{above1}
\begin{array}{cc}
dH_{\alpha}^{'}=\{H_\alpha^{'}, p_0+H_0\}d\tau + \{H_\alpha^{'},H_\beta^{'} \}dt_{\beta},& \alpha, \beta=1,\cdots K.\\
\end{array}, \ee
$H_0=H_c$ There are two cases to be studied. First, all of the
variations (\ref{above1})might vanish identically. In this case
there is no need to go one further step. The system is
integrable.Second, some of the variations might vanish
identically, and the other variations, say m of them, do not
vanish. These non-vanishing variations will give (m) differential
equations in $\dot t_{\beta}= {dt_\beta\over d\tau}$.
 Here one should notice
that although we impose the conditions that $t_\beta$ are
independent variables, theory forces us to calculate them as the
solutions of differential equations which arise from the
variations. Now, since all variations are zero the system is
integrable. The price we pay for this is the determination all of
the independent variables as function of $\tau$. As we can see
from (\ref{above1}), when $\tau=t$, $\dot t_{\beta}$ is nothing
that the expression of the Lagrange multiplier from chain
corresponding to the primary constraint $H_\beta$. This result is
valid in general , at the final stage , all velocities of the
gauge variables are the Lagrange multipliers of the chain method.
 The next step in the (HJ) formalism is to calculate the action.
 Since the system is second-class, the "Hamiltonians" are not in
 involution. One way to solve this problem is to try to solve all equations of motion.When we are dealing
 with non-linear equations this way is very difficult to be used.
Another way to bypass this problem is
 to make the "Hamiltonians" in involution by  selecting half of
 them and
 transforming $H_0^{'}$ as in  (\ref{extra}) for discrete case
 or as in (\ref{extra1}) for fields.

\subsection{Examples}

 To illustrate the approaches we will
 present two examples.

 {\bf 1.}Consider the Lagrangian \be L={1\over 2}({\dot q_1}+q_5)^2+{1\over
2}({\dot q_2}+q_6)^2 + {1\over 2}({\dot q_3}^2+{\dot
q_4}^2)-q_5(q_2+V_1(q_3,q_4))+q_6(q_1+V_2(q_3,q_4))-V_3(q_3,q_4),
 \ee
 where $V_1, V_2$ and $V_3$ are any functions of their arguments.
Let us apply (HJ) formalism for this system. The primary
constraints are

\be\label{integrab} H_1^{'}\equiv p_5, H_2^{'}\equiv p_6
 \ee

 The usual canonical Hamiltonian is
 \be
 H_0={1\over
 2}\sum_{i=1}^{4}{p_i}^2-q_5(p_1-q_2-V_1(q_3,q_4))-q_6(p_2+q_1+V_2(q_3,q_4))+V_3(q_3,q_4),
 \ee
 so
 \be\label{integrabh}
 H_0^{'}=p_0 +H_0
 \ee
 These expressions give the
 following equations of motion:
 \be\label{ecuaaa}
 \begin{array}{ccc}
 dq_1=(p_1-q_5)dt,&  dq_2=(p_2-q_6)dt, dq_3=p_3dt,dq_5=dq_5,  \\
 dq_4=p_4dt,& dp_1= q_6dt, dp_2=-q_5 dt,dq_6=dq_6\\
 dp_3=(-q_5{\partial V_1\over\partial
 q_3}+ q_6{\partial V_2\over\partial q_3}-{\partial V_3\over\partial
 q_3})dt & dp_4=(-q_5{\partial V_1\over\partial
 q_4}+ q_6{\partial V_2\over\partial q_4}-{\partial V_3\over\partial
 q_4})dt.
 \end{array}
 \ee

 Using (\ref{ecuaaa}) variations of constraints $H_1^{'}$ and
 $H_2^{'}$ follows as

\be dH_1^{'}=\{H_0^{'},H_1^{'}\}dt
+\{H_2^{'},H_1^{'}\}dq_6,\\dH_2^{'}=\{H_0^{'},H_2^{'}\}dt
+\{H_1^{'},H_2^{'}\}dq_5,\ee so
 \be\label{integrab1}
 H_3^{'}=p_1-q_2-V_1, H_4^{'}=p_2+q_1+V_2.
 \ee

Imposing the variations of $H_3^{'}$ and $H_4^{'}$ to be zero and
taking into account (\ref{integrab1}) we obtain another two new
"Hamiltonians":

\be\label{final} H_{5}^{'}=2q_6-p_2-{\partial V_1\over\partial
q_3}p_3 -{\partial V_1\over\partial q_4}p_4 , H_{6}^{'}=-2q_5 +p_1
+{{\partial V_2}\over\partial q_3}p_3+{\partial V_2\over\partial
q_4}p_4. \ee If we consider the variations of (\ref{final}) we
obtain a first order equation for $q_5$ and $q_6$. A simple
calculation shows that ${\dot q_5}$ and $\dot q_6$ have the same
form as the Lagrange multipliers from chain method. At this point
we conclude that our results are in agreement to \cite{mitra} .
  If we choose, for example,
$V_1(q_3,q_4)=q_3,V_2(q_3,q_4)=q_4$ and $V_3(q_3,q_4)=0$, then
(\ref{ecuaaa}) becomes \ba\label{hehe}
&dq_1=&(p_1-q_5)dt,dq_2=(p_2-q_6)dt, dq_3=p_3dt,\\
&dq_4=p_4dt&,dp_1=q_6dt,dp_2=-q_5dt,dp_3=-q_5dt,dp_4=q_6dt.\ea The
solution of (\ref{hehe}) is

\ba  &q_6(t)&= C_9\cos(t)-C_{10}\sin(t), q_5(t)=
C_9\sin(t)+C_{10}\cos(t),\\
 &q_4(t)&=
-C_9\cos(t)+C_{10}\sin(t)+C_7t+C_5,\\
&q_3(t)&= C_9\sin(t)+C_{10}\cos(t)+C_8t+C_6, q_2(t) = C_4t+C_1,\\
&q_1(t)&= C_3 t+C_2, p_1(t) = C_9\sin(t)+C_{10}\cos(t)+C_3,\\
&p_2(t)&= C_9\cos(t)-C_{10}\sin(t)+C_4,p_3(t) =
C_9\cos(t)-C_{10}\sin(t)+C_8,\\
&p_4(t)& = C_9\sin(t)+C_{10}\cos(t)+C_7,
 \ea
 where $C_i,i=1,\cdots 10$ are constants.

 Imposing $H_3^{'}=H_4^{'}=H_5^{'}=H_6^{'}=0$ we obtain the
 following restrictions on the above constants
  \be
 C_3=C_7, C_4=-C_8, C_4=-C_2-C_5, C_3=C_1+C_6.
 \ee

  As it is seen the above example is two-chain example and three
constraints are commuting each other. The main aim was to make the
system integrable and to find the action.One way is to introduce
the expressions of $q_i,i=1\cdots 6$ in $H_0^{'}$ and then we will
find the action using only one "Hamiltonian". In the following we
will apply the other method, mainly we will use only half of the
constraints and we will try to modify the Hamiltonian $H_0^{'}$
such that it will be involution with them. Having in mind to
obtain an integrable system and with physical interpretation from
(HJ) point of view we choose the following "Hamiltonians"

\be H_1^{'}=p_5, H_2^{'}=p_6, H_3^{'}=p_1-q_2-V_1 \ee

Using (\ref{extra}) the new form of $H_0^{'}$ is

\be\label{sus1}\begin{array}{cc}
 H_0^{''}= p_0+{1\over 2}(p_1^2+p_2^2 +p_4^2) +{1\over
8}(p_2-q_1-V_2)^2 +V_3&\\
 -{1\over
2}(p_2+q_1+V_2)(2q_6-p_2-{\partial V_1\over\partial  q_3}p_3-
{\partial V_1\over\partial q_4}p_4) + {1\over 8}(p_2+q_1+V_2)^2(1
+({\partial V_1\over\partial q_3})^2 +({\partial V_1\over\partial
q_4})^2)
\end{array} \ee

Since $H_1^{'}, H_2^{'}, H_0^{''}$ are commuting the corresponding
system is integrable.

The corresponding action is

\be S=\int {dt\{{1\over 2} (p_3^2+p_4^2)+{1\over
8}[{p_2}^2-(q_1+V_2)^2][1+({\partial V_1\over\partial q_3})^2
+({\partial V_1\over\partial q_4})^2]\}} \ee
 {\bf 2.} The bosonized Lagrangian for
the chiral model in (1+1) dimensions is written as \cite{fad}

 \be L=\int dx[{1\over
2}(\partial_{\mu}\phi+A_{\mu})(\partial^{\mu}\phi
+A^{\mu})-{1\over
4}F_{\mu\nu}F^{\mu\nu}-\varepsilon^{\mu\nu}{\partial_{\mu}\phi}A_{\nu}]
 \ee
The canonical Hamiltonian density becomes

\be\label{haa} H_c=\int dx[{1\over 2}(\pi^2+(\partial_x
\phi)^2+\pi_1^2) +\pi_1{\partial_x A_0}+(\pi +\partial_x \phi
+A_1)(A_1-A_0)],
 \ee
 where $\pi,\pi_0, \pi_1$ represent the momenta conjugate to
 $\phi,A_0,A_1$ respectively. In (HJ) formalism the primary Hamiltonian density is

\be\label{primatv} H_1^{'}=\pi_0.\ee The equations of motions
corresponding to (\ref{haa}) and (\ref{primatv}) are

\be\label{coco}
 \begin{array}{c} d\phi=(\pi+A_1-A_0)dt, dA_1=(\pi_1+{\partial_x
A_0})dt,\\
d\pi=[\partial_x(A_1-A_0)+
{\partial_x^{2}}\phi]dt,d\pi_1=-(2A_1+A_0-\phi-{\partial_x}\phi)dt
\end{array}
 \ee

 Taking into
account the variation of (\ref{primatv}) we obtain the second
"Hamiltonian" as \be\label{prima1} H_2^{'}={\partial _x
(\pi_1+\phi)}+\pi +A_1 \ee From (\ref{prima1}) we obtain another
"Hamiltonian"
 density as
\be\label{prima2} H_3^{'}=\pi_1 \ee Finally, imposing the
variation of $H_3^{'}$ to be zero we obtain the last "Hamiltonian"
density as \be H_4^{'}=-\pi-\partial_x\phi-2A_1+A_0 \ee We
identified the matrix $M_{ij}$ from chain method (\ref{matrice2})
as

 \be\label{matrice1} M(x,y)=\left(\begin{array}{cccc}
0 & 0 & 0  & -1\\
0 & 0 & 1  &  0 \\
0 & -1 & 0 & 2  \\
1 & 0 &  -2& 0
\end{array}\right)\delta(x-y).
\ee

As we can easily  see from (\ref{matrice1}) the "Hamiltonians"
$H_1^{'}, H_2^{'}, H_3^{'}, H^{'}$ are not in involution and using
(\ref{condh})the system of equations (\ref{coco})is not
integrable. In addition, we mention that if we continue to
consider the variation of $H_4^{'}$ we will obtain that ${\dot
A_0}$ is nothing the value of the Lagrange multiplier in the
Dirac's formulation.
  The modified canonical density Hamiltonian becomes

 \be
H_0^{''}=p_0+H_c-{1\over 2}\pi_1^2. \ee

We have three "Hamiltonian" densities which are in involution

\be\label{action}
\begin{array}{c}
H_0^{''}=p_0+{1\over 2}(\pi^2+(\partial_x \phi)^2)
+\pi_1{\partial_x A_0}+(\pi +\partial_x \phi +A_1)(A_1-A_0),\\
H_1^{'}=\pi_0, H_2^{'}=\pi+{\partial _x (\pi_1+\phi)} +A_1
\end{array} \ee

If we ignore the surface term, the action corresponding to
(\ref{action}) becomes \be S=\int{dx dt}[-{1\over
2}(\pi^2+(\partial_x \phi)^2)- (\pi +\partial_x \phi
+A_1)(A_1-A_0)-{\dot\phi}A_1]. \ee

\section{Conclusions}

 In this paper we proved that the chain method and (HJ) formalism are
equivalent. In other words both stabilization algorithms gave the
same set of constraints. Since the system is second-class
constrained it appears that the last Hamiltonian or Hamiltonian
density in (HJ) contains a variable which was undefined by the
system of the equations corresponding to the primary constraints
and the Hamiltonian $H_0^{'}$. If we take the variation of the
last Hamiltonain we obtained the velocity of that variable. We
have shown that this velocity is nothing that the Lagrange
multiplier in  Dirac's formalism. In fact, in (HJ) formalism we
have no Lagrange multipliers to start with but we have a set of
variables $q_\alpha$ corresponding to the primary constraints
$H_{\alpha}^{'}=p_{\alpha}+H_{\alpha}$, $\alpha = 1,\cdots K $.
And the end of the stabilization algorithm of (HJ) we obtained a
set of "Hamiltonians" and a set of new equations for $\dot
q_{\alpha}$. The remaining problem is to analyzed the
integrability of the system of total differential equations. For
second class-constrained systems (HJ) formalism is not integrable
in the sense of (\ref{condh}). To make it integrable we have
several options corresponding to several techniques of modifying
the set of second class-constraints into a first-class one. For
(HJ) formalism this step is crucial because the action delivered
by it depends drastically of the form of the "Hamiltonians". In
other words only for  "Hamiltonians" in the form
$H_{\alpha}^{'}=p_{\alpha}+H_{\alpha}$ it is possible to calculate
the action given by (\ref{(z)}). Following this idea we calculated
 the action of (HJ) corresponding to the gauge system obtained after
 applying the chain method.

 \section {Acknowledgments}
 One of the authors (D. B.) would like to thank M. Henneaux for
 stimulating discussions and encouragements.
This work is partially supported by the Scientific and Technical
Research Council of Turkey.

\end{document}